\documentclass[runningheads]{llncs}

\usepackage[T1]{fontenc}
\usepackage{newtxtext}
\usepackage[varvw]{newtxmath}
\usepackage{graphicx}
\usepackage{color}
\usepackage{hyperref}
\hypersetup{
    colorlinks=true,
    linkcolor=blue,
    citecolor=blue,
    urlcolor=blue
}
\begin{document}
\title{Computational Raga Recognition: A Theory-Driven Approach }

\author{Janay Sukheja \and Shailaja D Sharma} 

\institute{ Azim Premji University, Bengaluru, Karnataka, India }
\maketitle
\begin{abstract}
Raga recognition in Hindustani classical music is difficult because a raga is not defined only by the notes used in it. Its identity also depends on the way notes are emphasized, the movement between notes and characteristic phrases. In this work, we develop a computational approach to analyse these characteristics directly from audio recordings. The method uses pitch extracted from recordings and represents it using a 44-position note framework based on the 22-shruti system. From this representation, Raga features based on the Indian classical musicology are extracted.
Each feature is compared with a reference representation of the raga and the resulting scores are combined to obtain an overall similarity score. The method is first studied using Raga Yaman and is then tested on recordings of other ragas. The computational results are also compared with the theoretical description of Yaman to identify which raga characteristics are captured well and where the method has limitations.
\end{abstract}
\begin{keywords}
Raga recognition, Hindustani classical music, computational musicology, raga feature extraction, pitch representation, 22-shruti framework, melodic phrase analysis, raga recognition model.
\end{keywords}
\section{Introduction}
\label{sec:introduction}
Raga recognition is an important problem in computational analysis of Hindustani classical music. Unlike a classification based only on the notes present in a recording, identifying a raga requires considering how those notes are used. Characteristics such as Vadi, Samvadi, Nyasa, characteristic melodic phrases and patterns of ascent and descent contribute to raga identity\footnote{See Appendix \ref{app:characterstics} for a description of Raga Characterstics}.This becomes particularly important for ragas that share similar pitch material. In this work, we investigate whether these theoretical characteristics can be extracted from audio recordings and used for computational raga comparison. We develop a feature-based framework in which different characteristics of a raga are extracted separately from pitch sequences and compared with a reference representation. Raga Yaman\footnote{See Appendix \ref{app: Yaman_Characterstics.} for description of Raga Yaman} is used as the primary case study. The resulting model is evaluated using recordings of Yaman as well as other distinct ragas, namely Shudh Kalyan, Bhoop, Bageshri, Miyan Malhar and Marwa.
The evaluation has two parts. First, the characteristics extracted from the recordings are compared with the theoretical description of Yaman. Second, the Yaman reference model is tested on recordings of different ragas to examine whether the similarity score is specific to Yaman. This allows us to evaluate both the correspondence between the computational representation and raga theory and its usefulness for raga recognition.

\section{RELATED WORK}
Computational raga recognition has been approached using pitch distributions, sequential representations, melodic phrases,complexity-based approaches and supervised learning methods\cite{b1}\cite{b2}\cite{b3}\cite{b4}\cite{b5}\cite{b6}. Pitch distribution approaches describe the relative occurrence of pitch classes or pitch regions in a performance and have been used to characterize ragas from their melodic content\cite{b1}. Sequence-based methods\cite{b6} retain the order of notes and can therefore capture transitions and other temporal relationships that are lost in a pitch distribution. Phrase-based approaches focus on recurring melodic patterns, which are particularly relevant to Hindustani music where characteristic phrases form an important part of raga identity\cite{b4}. Another approach uses Compression-Complexity Measures based on Lempel-Ziv complexity to investigate structural relationships between Carnatic music compositions, particularly between Melakarta and Janya ragas\cite{b5}. Other approaches use statistical and machine learning models to combine different musical features for raga classification. Hidden Markov Models, Support Vector Machines and other classifiers have been applied to pitch and temporal representations \cite{b2}, while more recent approaches have explored learned representations for automatic raga recognition. These approaches motivate the use of multiple types of melodic information, but the present work takes the analysis in a different direction. Instead of using general purpose features to directly learn a raga class, we use characteristics that are explicitly defined in the description of a raga. The aim is to retain a direct connection between each computational feature and its corresponding musicological concept. This also allows the individual characteristics to be examined separately when comparing a recording with a raga reference.
In this work, these characteristics are combined into a computational framework and evaluated using Raga Yaman as a case study.

\section{Methodology}
The proposed methodology provides a computational framework for identifying raga characteristics from Hindustani classical vocal recordings. The overall process consists of audio pre-processing, pitch extraction, tonic normalization, pitch representation, raga feature extraction and similarity calculation. The extracted pitch sequence is represented using an extended 44-position framework based on the 22-shruti system. Six important raga characteristics, namely Ny\={a}sa, Purv\={a}ng–Uttar\={a}ng, Vadi–Samvadi, Alpatva/Varjit, Aarohan–Avrohan and Chalan, are then extracted from the pitch representation. Each characteristic is compared with the corresponding Yaman reference model and the individual similarity scores are combined to obtain the overall raga similarity score. The complete methodological framework is shown in Figure 1.

\begin{figure}[!htb]
    \centering
    \setlength{\fboxrule}{0.8pt}   
    \setlength{\fboxsep}{3pt}      
    \fbox{\includegraphics[width=\dimexpr\textwidth-2\fboxrule-2\fboxsep\relax]{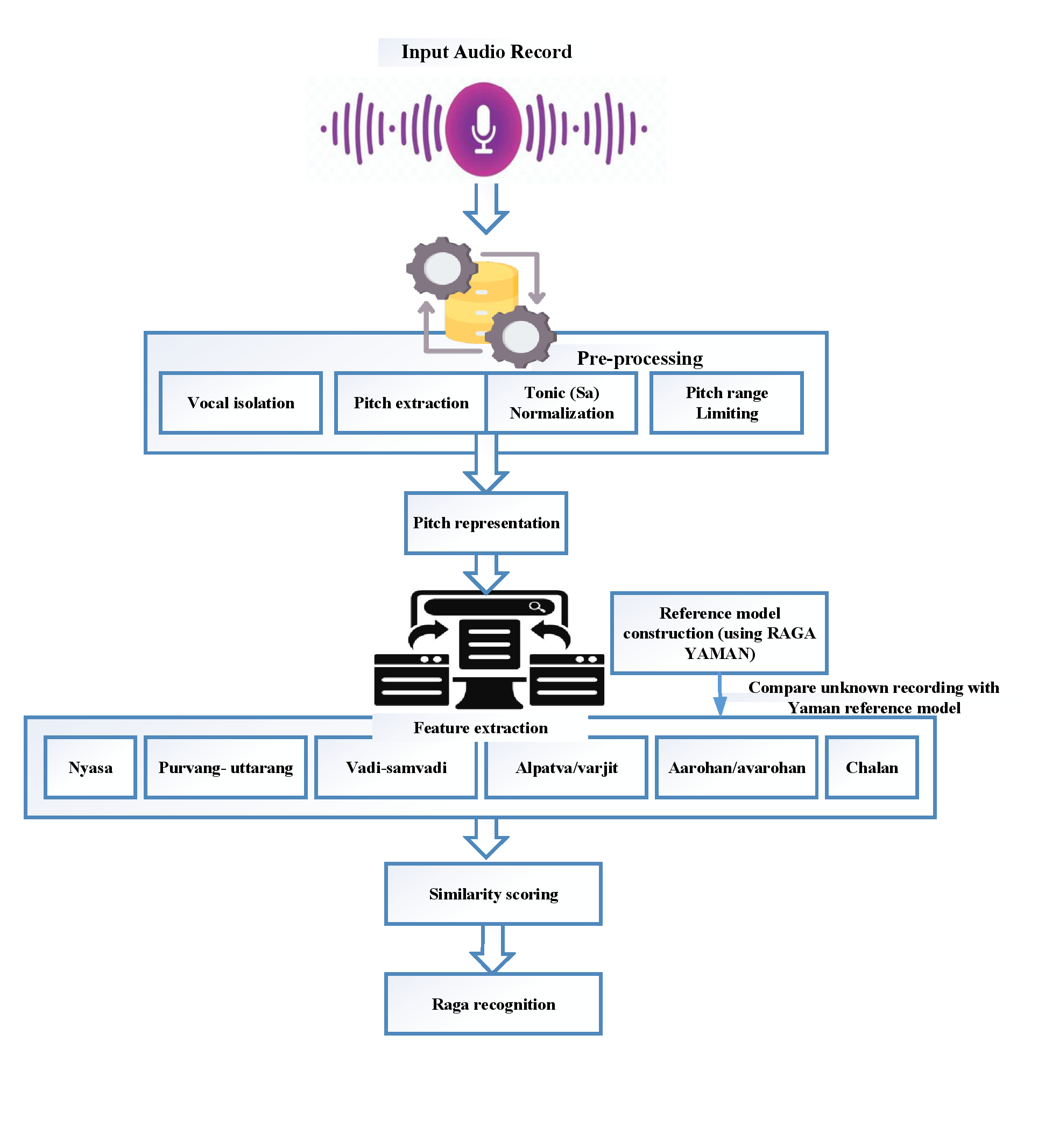}}
    \caption{Complete methodological framework}
    \Description{Flowchart illustrating the complete methodological framework.}
    \label{methodology_diagram}
\end{figure}
\subsection{DATA COLLECTION AND PREPROCESSING}
The analysis was carried out on recordings of Hindustani classical performances. Raga Yaman was used as the primary raga for developing the reference representation. 11 audio recordings of approximately 6 hours total were collected from performances by expert musicians. The lead portions were isolated before pitch extraction using RipX daw to reduce the influence of accompaniment and other instruments. Pitch was extracted from the audio using the pYIN algorithm implemented in Librosa\cite{b9}\cite{b10}. The audio was sampled at 44.1 kHz and the pitch contour was initially extracted at a temporal resolution of 10 ms. The pitch range was restricted from C1 (32.70 Hz) to C6 (1046.50 Hz). The tonic Sa was determined manually for each recording. The extracted pitch values were then represented relative to the tonic and mapped to the 44-position note framework described in the following Section. This representation was used for the extraction of the different raga characteristics.
For most features, the pitch sequence was retained at the 10 ms resolution. For Chalan analysis, the sequence was further converted to 250 ms intervals.

\subsection{PITCH REPRESENTATION}
The pitch representation used in this work is based on the 22-shruti framework.  Refer to ~\ref{app:shruti}

\begin{figure}[htb!]
    \centering
    \setlength{\fboxrule}{0.8pt}   
    \setlength{\fboxsep}{2pt}      
 \fbox{\includegraphics[width=1.0\linewidth]{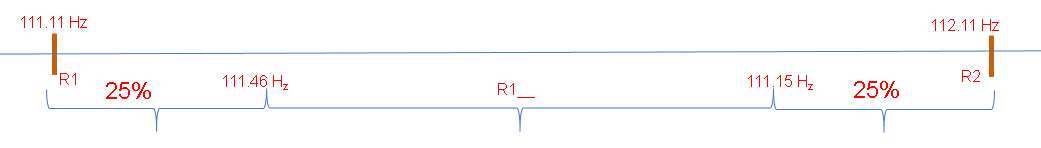}}
    \caption{Visual representation of intermediate note}
    \label{fig:Pitch_rep}
\end{figure}

Since the extracted pitch does not always fall exactly on one of the theoretical shruti positions, the representation was extended to 44-note positions by including the intermediate positions between the 22 shrutis based on a partioning of the inter-shruti interval in a 25\%-50\%-25\% configration as shown in Fig\ref{fig:Pitch_rep}. This provides a finer representation of pitch movement while retaining the theoretical shruti structure.
For each extracted pitch value, its position relative to the tonic was calculated and mapped to the nearest position in the 44-note framework. The same representation was used across the different feature extraction procedures, allowing note distributions and note sequences to be compared between recordings.

\subsection{RAGA FEATURE EXTRACTION}
The mapped note sequences were used to extract six raga characteristics(see ~\ref{app:characterstics}): Nyasa, Purvang-Uttarang, Vadi-Samvadi, Alpatva/Varjit, Aarohan-Avrohan and Chalan. All features except Chalan were analysed using the 10 ms note sequence, while Chalan was analysed separately using the 250 ms sequence.

\subsubsection{NYASA AND PURVANG-UTTARANG}
\label{Nyasa_extraction}
Nyasa swaras were identified from sustained notes in the 10 ms sequence. A note that remained stable for at least 100 consecutive frames, corresponding to approximately 1 second, was treated as a Nyasa occurrence. The frequency of these occurrences was calculated for each position in the 44-note representation.
For Purvang-Uttarang analysis, octave information was retained with the mapped note sequence. The modal Sa octave (Madhya-Sthayi) was used as the reference octave. The notes in each selected octave were divided into Purvang and Uttarang according to their position relative to Sa and their relative frequencies were calculated for each recording.

\subsubsection{VADI-SAMVADI AND ALPATVA/VARJIT}
Vadi and Samvadi were determined from the Nyasa distribution, Purvang-Uttarang emphasis and overall note prominence. The most prominent Nyasa swara in the dominant region was selected as the Vadi candidate, while the corresponding prominent swara in the complementary region was selected as the Samvadi candidate.
For note prominence, the six most frequently occurring note positions were considered rather than only the single most frequent position. This reduces the effect of pitch variation and ornamentation, which can spread a prominent swara across neighbouring positions in the 44-note representation.
The same six-note set was used for Alpatva and Varjit analysis. Notes expected to occur sparingly or to be avoided in the reference raga were treated as deviations when they appeared among the dominant notes.

\subsubsection{AAROHAN-AVROHAN}
Aarohan and Avrohan were represented using transitions between stable swaras. A note that remained unchanged for at least 0.75 seconds was treated as a stable note. Subsequent stable notes within a 5-second window were examined and the resulting transitions were separated into ascending and descending movements.

\subsubsection{CHALAN}
Chalan was analysed as a phrase-level feature using the 250 ms note sequence. Overlapping phrases of 4–7 seconds were extracted using a sliding window, giving 16–28 note positions per phrase.
Because the same melodic movement can occur with small variations between performances, exact matching was not used \cite{b6}. Instead, phrases were compared using a weighted  Ratio ~\cite{b7} \cite{b8} ratio in which substitution costs depended on the distance between the corresponding positions in the 44-note framework. A similarity threshold of 0.9 was used to group similar phrases. The phrase groups were ranked by frequency and the most frequent groups were taken as candidate Chalan patterns. The top ten representatives were retained for each phrase duration in the reference model.

\subsection{SIMILARITY AND SCORING}
For each feature, a reference model was constructed from the Yaman recordings. The corresponding feature extracted from a test recording was then compared with the reference model using the feature-specific similarity criterion described below.
\subsubsection{NYASA SIMILARITY}

The Nyasa distribution of each recording was represented as a 44-dimensional vector.

\[
N = (N_1, N_2, \ldots, N_{44})
\]

where $N_i$ denotes the percentage occurrence of the $i$-th note position among the detected Nyasa swaras. The reference vectors were normalised to unit length and averaged to obtain the mean reference directional vector $N_{Ref}$.

The angular distance between all 11 test vectors $N$ and the reference vector $N_{Ref}$ was calculated. The median\footnote{Median was chosen for its relative stability compare to mean} angular distance and one standard deviation were used to determine an acceptance threshold, allowing for the natural variation present among different performances. A new test recording $N_{new}$ was considered consistent with the reference Nyasa distribution if the angular distance of $N_{new}$ and $N_{Ref}$ is within the threshold.

\subsubsection{PURVANG-UTTARANG SIMILARITY}
The Purvang-Uttarang representation was compared using the relative occurrence of notes in the two regions. Let
\[
P=(P,U)
\]
where P and U denote the Purvang and Uttarang proportions, respectively. The reference vectors were first normalised to unit length. Reference recordings was incorporated by identifying the pair of reference vectors with the maximum angular separation. These vectors represent the extreme boundaries of the Purvang/Uttarang distribution observed within the reference dataset. The median directional vector was computed. The angular distances between the median direction and each of the two extreme reference vectors were used to define the acceptance region with additional 5\%  tolerance margin. The angular distance between the reference vector and test vector was measured.
A test recording falling within this region was considered consistent with the reference Purvang-Uttarang distribution.

\subsubsection{VADI-SAMVADI SIMILARITY}
The six most frequent note positions obtained during feature extraction were used to evaluate Vadi-Samvadi prominence. Let
\[
D_6 = \{d_1, d_2, \ldots, d_6\}
\]
denote the six dominant positions. The presence of the theoretical Vadi and Samvadi within $D_6$ was used to determine their agreement with the reference raga.

The contribution based on the Nyasa distribution and Purvang-Uttarang region was evaluated separately through their respective similarity scores and was therefore not counted again in this component.

\subsubsection{ALPATVA/VARJIT SIMILARITY}
The same dominant note set $D_6$ was compared with the expected Alpatva and Varjit swaras of the reference raga. The presence of a swara expected to occur sparingly or to be avoided was treated as a deviation and contributed a penalty to the overall score.

\subsubsection{AAROHAN-Avrohan SIMILARITY}

For each observed transition from swara $i$ to swara $j$, the transition probability was calculated as:
\[
I_{ij} = \frac{N(i \rightarrow j)}{\sum_k N(i \rightarrow k)}
\]
where $N(i \rightarrow j)$ denotes the number of observed transitions from $i$ to $j$.

The reference model was filtered by removing transitions occurring fewer than five times or having a probability below 0.1.

The three most probable remaining transitions for each starting swara were retained. Transitions observed in the test recording were then compared with this reference set. A matched transition contributed its reference probability to the similarity score (S) 
where $M$ is the set of matched transitions and $T_{ij}^{Ref}$ is the corresponding reference transitional probability.

\[
S = \sum_{(i,j)\in M} T_{ij}^{Ref}
\]

\subsubsection{CHALAN SIMILARITY}
The Chalan representatives obtained during feature extraction were compared with the corresponding reference representatives using the weighted Levenshtein Ratio \cite{b7}. In the notation below, X is the test sequence and R is the refrence sequence. $L_W (X,R)$ refers to the weighted Levenshtein Ratio. A test reference phrase pair was counted as a match when
\[
L_W (X,R) \geq 0.9
\]
The number of matched phrases across the four durations, 4–7 seconds, was then converted into the Chalan score.

\section{Results}
The results were evaluated in two stages. First, the characteristics obtained from the Yaman reference recordings were examined to see how consistently they appeared across performances. These results were then compared with the theoretical description of Yaman (see \ref{app: Yaman_Characterstics.}). Second, the reference model was tested on recordings of other ragas to examine whether the resulting similarity scores distinguished them from Yaman.

\subsection{NYASA ANALYSIS}
\begin{figure}
    \centering
    \includegraphics[width=1.00\linewidth]{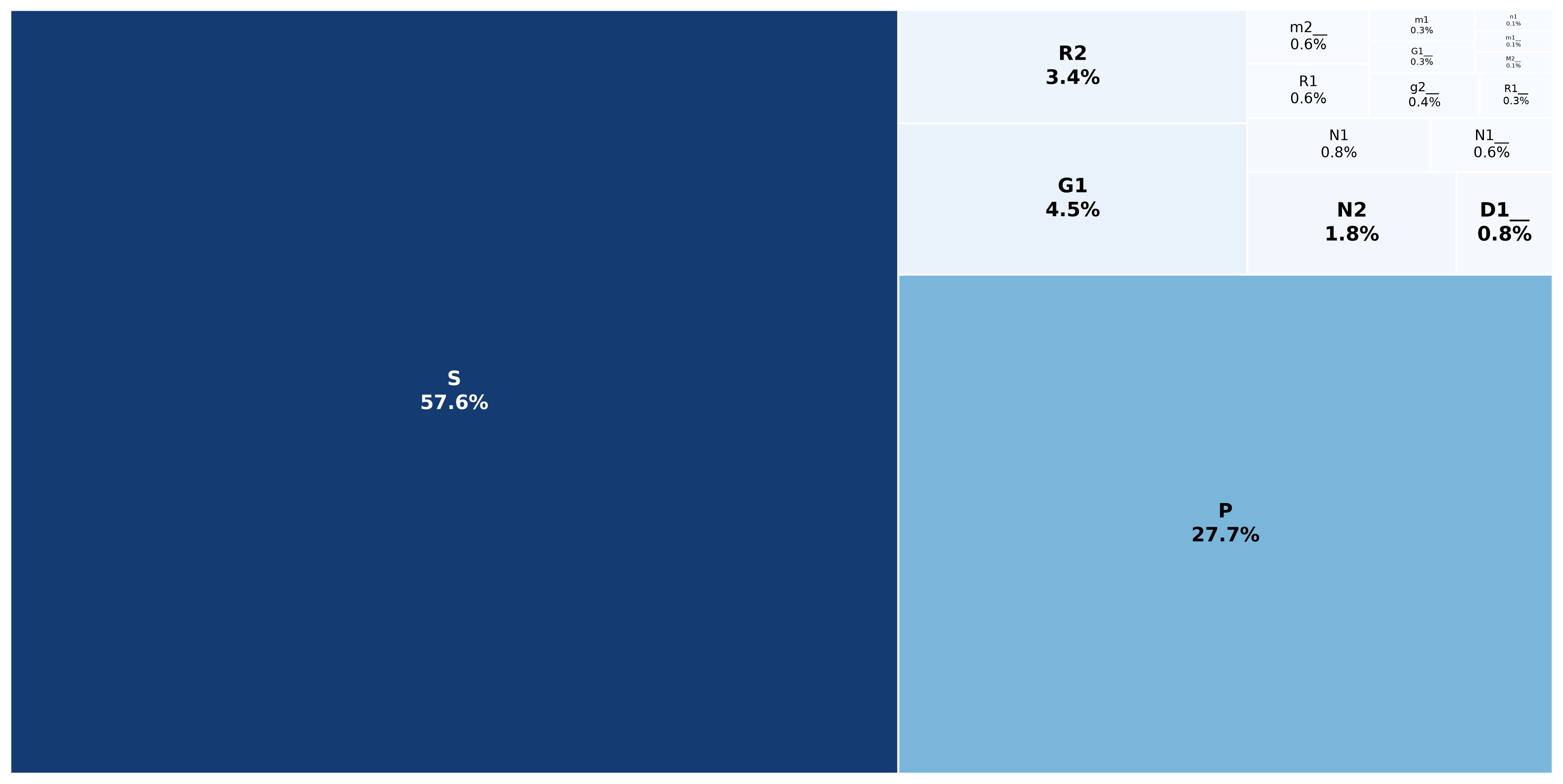}
    \caption {Nyasaswara distribution in reference recording}
    \label{fig:Nyasa}
\end{figure}

Ny\={a}sa occurrences were extracted from the Yaman reference recordings using the 1-second stability criterion described in Section\ref{Nyasa_extraction}. A total of 714 Nyasa occurrences were detected across the reference recordings. Sa and Pa were the most frequent Ny\={a}sa positions, with 411 (57.56\%) and 198 (27.73\%) occurrences, respectively. Among the remaining swaras, Ga, Re and Ni were the most prominent Ny\={a}sa swaras refer to Fig\ref{fig:Nyasa}.

\subsection{PURVANG-UTTARANG ANALYSIS}
\begin{figure}[htb!]
    \centering
    \includegraphics[width=1.00\linewidth]{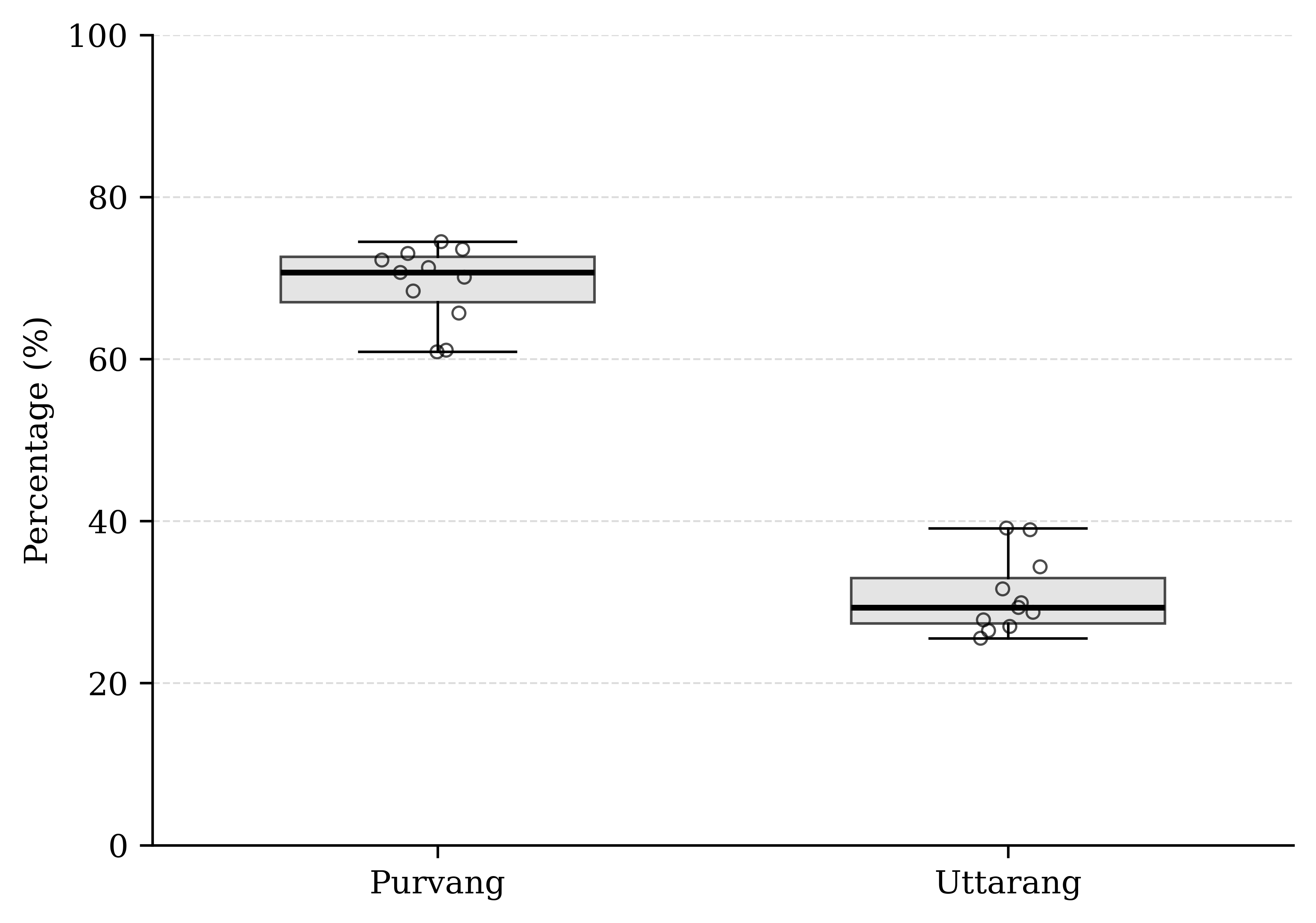}
    \caption{Purvang/Uttarang distribution in reference recording}
    \label{fig:Uttarang}
\end{figure}

The Purvang-Uttarang distribution was calculated for each Yaman recording after identifing the modal octave as the Madhya Saptak. The proportion of notes falling in the Purvang and Uttarang regions was then calculated.
Across the eleven Yaman recordings, the Purvang proportion ranged from 60.87\% to 74.47\%, while the corresponding Uttarang proportion ranged from 25.53\% to 39.13\% as shown in Fig\ref{fig:Uttarang}. The mean proportions across the recordings were 69.11\% for Purvang and 30.89\% for Uttarang.

\subsection{VADI-SAMVADI ANALYSIS}

The Vadi-Samvadi candidates were determined from the Nyasa distribution, Purvang-Uttarang distribution and overall note frequency. Using these criteria, S and P emerged as the most prominent candidate pair, followed by G and N. The present implementation, however, does not account for melodic context, such as the occurrence of a swara at the end of a melodic phrase, which is relevant to Vadi-Samvadi identification. Thus, the two candidate pairs obtained empirically were S-P and G-N.

\subsection{ALPATVA/VARJIT ANALYSIS}
For the Alpatva/Varjit analysis, the note positions in the Yaman reference recordings were ranked according to their overall frequency. The six most frequent positions were retained as the dominant reference set. These were S, N, G, P, R, m. The remaining note positions occurred less frequently in the reference recordings. Their occurrence in a test recording was therefore considered in relation to the dominant reference distribution when calculating the Alpatva/Varjit similarity. In particular, a note expected to occur sparsely or to be avoided in the theoretical description of the raga was treated as a deviation when it appeared among the dominant positions of a recording.

\subsection{AAROHAN-AVROHAN ANALYSIS}
\begin{figure}[htb!]
    \centering
    \setlength{\fboxrule}{0.8pt}
    \setlength{\fboxsep}{2pt}
    \fbox{\includegraphics[width=1.0\linewidth]{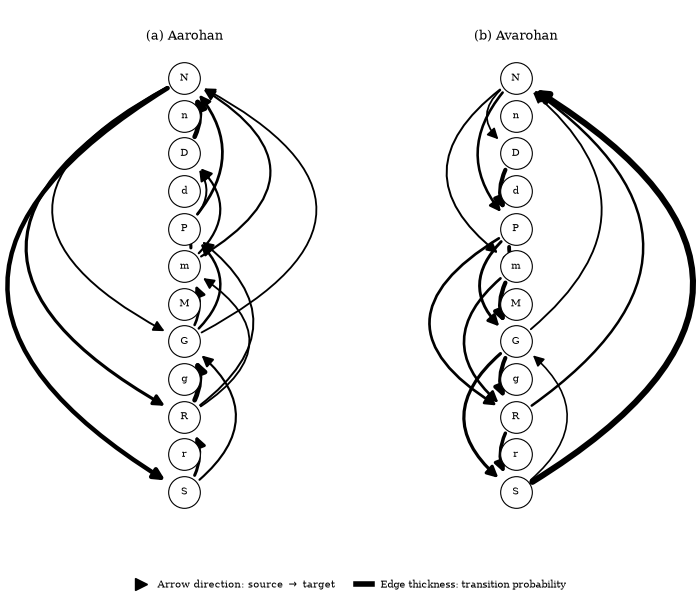}}
    \caption{Transition diagrams for Arohan and Avarohan.}
    \label{fig:arohan_avarohan}
\end{figure}

The transition structure obtained from the Yaman reference recordings is shown in Fig\ref{fig:arohan_avarohan}. The dominant transitions in the two directions were N→S in Aarohan and S→N in Avrohan. Other prominent movements included N→R, R→G and G→m in Aarohan, while m→G, D→P, G→R and P→m were prominent in Avrohan. Notably, N →R was more probable than S →R, consistent with the characteristic movement from N to R in Yaman.

\subsection{CHALAN}
Most of the characteristic melodic phrases described for Yaman were retained within the top-ten computationally extracted Chalan patterns. The characteristic melodic movements could be identified directly in these extracted sequences. For example, the theoretical N R S movement is represented in the extracted pattern N1 ••• R2 R2 N2 S S\footnote{The dots represent elongation of a note}. Similarly, the N D N movement occurs in the extracted pattern N1 ••• D1 n1 D2 n2 N1, while the M D N movement is represented by m2 ••• d2 n2 D1 D1 n2. These examples show that the characteristic melodic movements described for Yaman are retained in the computationally extracted Chalan patterns.

\subsection{RAGA RECOGNITION RESULTS}

The proposed scoring framework was evaluated using recordings of Raga Yaman and reference models constructed from the extracted musical features. The framework produced the highest overall score for Yaman, with a score of 84.0\%. The corresponding scores for Shuddha Kalyan, Bageshri, Miya Malhar and Marwa were 64.90\%, 18.90\%, 13.50\% and 0.0\%, respectively. The highest score for Yaman indicates that the combination of Nyasa, Purvang–Uttarang distribution, characteristic swaras, Aarohan–Avrohan transitions and Chalan provided substantial agreement with the reference model. Shuddha Kalyan produced the closest competing score, whereas the scores for Bageshri, Miya Malhar and Marwa were considerably lower. 

The results also showed that the individual features do not contribute equally across performances. In particular, the Nyasa component was more sensitive to performance variation, while characteristic Chalan patterns and Aarohan–Avrohan transitions were retained more or less consistently across the analyzed Yaman recordings. For reliable extraction of these characteristics, the input recording must contain sufficient musical material in which the relevant swaras and melodic movements are expressed. 

Recordings containing extended slow elaborations (alap, vilambit, or madhya laya compositions) provide more stable notes and characteristic melodic patterns for the feature extraction process. In contrast, recordings dominated by fast (drut) passages may contain insufficient sustained notes and characteristic melodic movements, making features such as Nyasa and Chalan more difficult to extract reliably. Thus, the effectiveness of the framework depends not only on the raga being performed but also on the musical content available in the input recording.

\section{DISCUSSION AND LIMITATIONS}

The results indicate that the proposed framework can distinguish Yaman from the other tested ragas using theoretically motivated musical features. The strongest agreement was obtained when the input recording contained sufficient musical material for the characteristic features of the raga to be observed. Recordings containing alap or slower and medium tempo passages provide greater opportunity for sustained notes and characteristic melodic movements to be captured, whereas recordings dominated by fast passages may provide insufficient evidence for features such as Nyasa and Chalan. The relatively high score obtained for the raga Shuddha Kalyan also confirms that some theoretical characteristics are shared between these ragas; therefore, the final score should be interpreted as a measure of agreement with the reference raga rather than as a probability of raga membership.

The Nyasa component was more sensitive to performance variation than the other features and may underestimate valid Yaman performances when characteristic resting notes are not sufficiently sustained or prominently expressed. The Vadi-Samvadi analysis also has a limitation: the current feature set determines candidates from note prominence but does not account for melodic context, such as the role of a swara at the end of a melodic phrase. Consequently, S-P and G-N emerged as candidate pairs. Improving the Nyasa model through a larger and more diverse reference dataset, together with incorporating melodic context into Vadi-Samvadi identification and investigating efficient similarity search methods such as Locality Sensitive Hashing (LSH), could improve the robustness of the framework. The current implementation also assumes that the tonic (Sa) is known beforehand and approximates it to the nearest Western pitch (e.g., C, C\#, or D), which can affect note mapping and subsequent feature extraction. Developing a more accurate tonic estimation method is therefore another important direction for future work.

\section{Conclusion}
This work developed a computational framework for raga recognition based on theoretically defined characteristics of Hindustani classical music. The empirical analysis of Yaman demonstrated that these characteristics can be computationally extracted and combined within a unified recognition framework, with the model assigning the highest score to the target raga. The results support the use of theory-driven musical descriptors as an alternative to approaches based primarily on learned representations. At the same time, the performance of the framework depends on the musical content available in the recording and on the completeness of the theoretical descriptors used. Further evaluation across a larger and more diverse set of ragas and performances is required to establish the broader applicability of the approach.

\section{AI USAGE STATEMENT}
Generative AI tools (ChatGPT, OpenAI) were used solely to improve the clarity, grammar and presentation of the manuscript. The research methodology, implementation, experimental design, results and conclusions were developed independently by the authors. The authors reviewed and verified all AI-assisted text and take full responsibility for the content of this manuscript.

\newpage

\appendix

\renewcommand{\thesection}{Appendix \Alph{section}}
\renewcommand{\thesubsection}{\Alph{section}.\arabic{subsection}}

\section{The 22 Shrutis}
\label{app:shruti}

The 22 shruti system used in this work refers from the work of Oak\cite{b11}. A shruti represents a specific pitch position within an octave. Unlike the Western equal-tempered system, where the frequency ratios between neighbouring notes are fixed, the positions of the shrutis in the Indian classical music system are described using specific frequency ratios.
The interval between neighbouring shrutis can be of three types: Poorna, Pramana and Nyuna. The frequency ratios for these intervals are
$$
\text{Poorna}=\frac{256}{243}, \qquad
\text{Pramana}=\frac{81}{80}, \qquad
\text{Nyuna}=\frac{25}{24}.
$$
These ratios are used to generate the relative positions of the 22 shrutis starting from Sa. The frequency of Sa is taken as the reference frequency. Pancham is obtained from Sa using the ratio

$$
P=\frac{3}{2}S
$$
where \(S\) is the frequency of Sa.
The remaining shruti positions are generated using the successive ratios in the sequence Poorna, Pramana, Nyuna, Pramana. In this way, the relative frequency of each of the 22 shrutis can be obtained with respect to the tonic Sa.
 
\noindent If \(S_0\) is the frequency of the tonic, the tonic in another octave is obtained using

$$
S_n=S_0 2^n
$$
where \(n\) represents the octave relative to the reference octave. The corresponding shruti frequencies in that octave are then obtained using the same ratios.

\begin{figure}[htb!]
    \centering
    \setlength{\fboxrule}{0.8pt}   
    \setlength{\fboxsep}{2pt}      
 \fbox{\includegraphics[width=0.4175\linewidth]{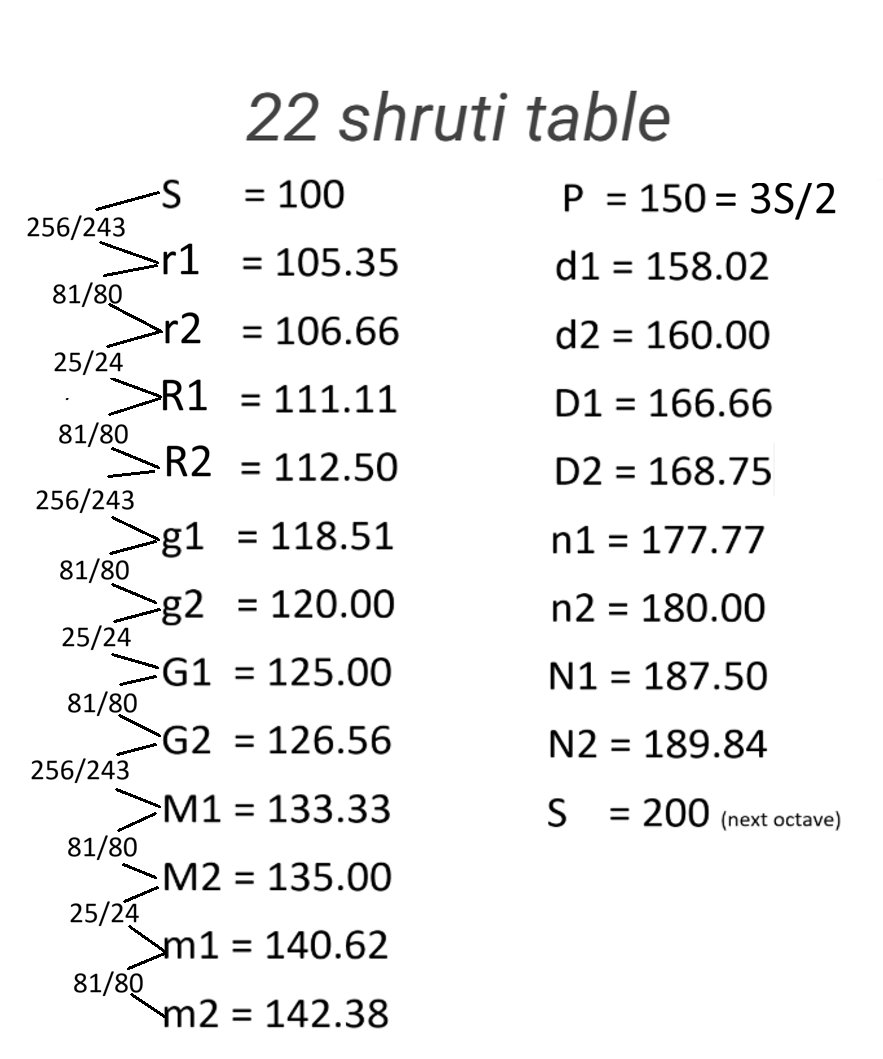}}
 \caption{22\_Shruti\_Table}\label{fig:Shruti_Table} \end{figure}

...
\newpage
\section{Raga Characterstics}
\label{app:characterstics}

A raga consists of many features \cite{b12} \cite{b13}. We group the features used in this paper into three categories for understanding the different aspects of a raga:

Spatial and Temporal Characteristics
Nyasa Swara
Purvang/Uttarang
Hierarchical Importance of Notes
Vadi
Samvadi
Alpatva/Varjit
Melodic Structure and Movement
Aarohan/Avrohan
Chalan

\subsection{Nyasa Swara}

Nyasa swaras are the resting notes in a raga where the performer tends to pause, sustain, or stabilize the melodic movement. A note which is repeatedly used as a resting point can have an important role in defining the character of a raga.

\subsection{Purvang/Uttarang}

In Indian Classical Music, the octave is commonly divided into two regions known as Purvang and Uttarang. Purvang consists of the lower region of the octave containing Sa, Re, Ga and Ma, while Uttarang consists of Pa, Dha, Ni and the upper Sa.
The relative use of these two regions can be different for different ragas.

\subsection{Vadi/Samvadi}

Vadi and Samvadi are considered the two most important swaras of a raga. The Vadi is regarded as the most prominent swara and has an important role in defining the identity of the raga. The Samvadi is the second most prominent swara and supports the role of the Vadi.
These swaras can be interpreted by looking at the notes that receive greater importance and emphasis in the raga. The Vadi is expected to have a stronger presence, while the Samvadi provides a complementary important swara.

\subsection{Alpatva/Varjit}

Alpatva and Varjit describe swaras that have limited or no use in a raga. An Alpatva swara is used sparingly, while a Varjit swara is avoided in the raga.

\subsection{Aarohan/Avrohan}

Aarohan and Avrohan describe the ascending and descending movement of a raga. Aarohan represents the characteristic movement of swaras in the ascending direction, while Avrohan represents the movement in the descending direction.
These movements describe how the swaras are connected to each other rather than only describing which swaras are present in the raga.

\subsection{Chalan}

Chalan represents the characteristic melodic movement of a raga. It consists of typical combinations and sequences of swaras that are commonly used while performing the raga.
Chalan can be interpreted through the recurring melodic phrases found in a raga.

\newpage

\section{Theoretical Characteristics of Raga Yaman}
\label{app: Yaman_Characterstics.}

Raga Yaman belongs to the Kalyan thaat (parent scale). It is generally classified as a Sampurna raga, using all seven swaras in both ascent and descent. It may also be rendered in an Audav--Sampurna form, with five notes in ascent and seven notes in descent.\cite{b14} \cite{b15}

\subsection{Swaras and Vadi--Samvadi}

Raga Yaman uses Teevra Madhyam ($m$) along with the Shuddha forms of the remaining swaras. Pancham ($P$) may be omitted in some phrases. The Vadi swara of the raga is Gandhar ($G$), while the Samvadi swara is Nishad ($N$).

\subsection{Aarohan and Avrohan}

The prescribed aarohan and Avrohan are:

\[
\text{Aaroh: } N\ R\ G\ m\ P\ D\ N\ S'
\]

\[
\text{Avrohan: } S'\ N\ D\ P\ m\ G\ R\ S
\]

The use of Nishad ($N$) in the aarohanis preferred and contributes to the characteristic expression of the raga.

\subsection{Key Phrases}

Some representative key phrases of Raga Yaman are:

\[
\begin{aligned}
&N R\ G;\quad R\ G;\quad N\ R\ S;\quad P\ G\ R;\quad m\ D\ P;\quad m\ D\ N
\end{aligned}
\]

These characteristic combinations, together with the prescribed swara structure and aaroh--Avrohan, contribute to the melodic identity of Raga Yaman.

\end{document}